\documentclass[onecolumn,showpacs,preprintnumbers,amsmath,amssymb]{revtex4}
\usepackage{graphicx}
\usepackage{dcolumn}
\usepackage{bm}
\begin{document}
\newcommand{\hs}{\hspace*{0.5cm}}
\newcommand{\vs}{\vspace*{0.5cm}}
\newcommand{\be}{\begin{equation}}
\newcommand{\ee}{\end{equation}}
\newcommand{\bea}{\begin{eqnarray}}
\newcommand{\eea}{\end{eqnarray}}
\newcommand{\ben}{\begin{enumerate}}
\newcommand{\een}{\end{enumerate}}
\newcommand{\bde}{\begin{widetext}}
\newcommand{\ede}{\end{widetext}}
\newcommand{\nn}{\nonumber}
\newcommand{\crn}{\nonumber \\}
\newcommand{\non}{\nonumber}
\newcommand{\noi}{\noindent}
\newcommand{\al}{\alpha}
\newcommand{\la}{\lambda}
\newcommand{\bet}{\beta}
\newcommand{\ga}{\gamma}
\newcommand{\va}{\varphi}
\newcommand{\om}{\omega}
\newcommand{\pa}{\partial}
\newcommand{\fr}{\frac}
\newcommand{\bc}{\begin{center}}
\newcommand{\ec}{\end{center}}
\newcommand{\Ga}{\Gamma}
\newcommand{\de}{\delta}
\newcommand{\De}{\Delta}
\newcommand{\ep}{\epsilon}
\newcommand{\varep}{\varepsilon}
\newcommand{\ka}{\kappa}
\newcommand{\La}{\Lambda}
\newcommand{\si}{\sigma}
\newcommand{\Si}{\Sigma}
\newcommand{\ta}{\tau}
\newcommand{\up}{\upsilon}
\newcommand{\Up}{\Upsilon}
\newcommand{\ze}{\zeta}
\newcommand{\ps}{\psi}
\newcommand{\Ps}{\Psi}
\newcommand{\ph}{\phi}
\newcommand{\vph}{\varphi}
\newcommand{\Ph}{\Phi}
\newcommand{\Om}{\Omega}
\def\lappeq{\mathrel{\rlap{\raise.5ex\hbox{$<$}}
{\lower.5ex\hbox{$\sim$}}}}


\title{Neutrino masses in the economical 3-3-1 model}

\author{P. V. Dong\footnote{On leave from Institute of
Physics, VAST, P.O. Box 429, Bo Ho, Hanoi 10000, Vietnam}}
\email{pvdong@iop.vast.ac.vn} \affiliation{Department of Physics and
NCTS, National Tsing Hua University, Hsinchu, Taiwan, R.O.C.}
\author{H. N. Long}
\email{hnlong@iop.vast.ac.vn} \affiliation{Department of Physics,
Kobe University, Nada, Kobe 657-8501, Japan} \affiliation{
 Institute of Physics, VAST, P.O. Box 429,
Bo Ho, Hanoi 10000, Vietnam\footnote{Permanent address} }
\author{D. V. Soa}
\email{dvsoa@assoc.iop.vast.ac.vn} \affiliation{Department of
Physics, Hanoi University of Education, Hanoi, Vietnam}

\date{\today}

\begin{abstract}

We show that, in frameworks of the economical 3-3-1 model, the
suitable pattern of neutrino  masses arises from the three quite
different sources - the lepton-number conserving, the spontaneous
lepton-number breaking and the explicit lepton-number violating,
widely ranging over the mass scales including the GUT one: $u\sim
O(1)\ \mathrm{GeV}$, $v\approx 246\ \mathrm{GeV}$, $\om\sim O(1)\
\mathrm{TeV}$ and $\mathcal{M}\sim \mathcal{O}(10^{16})\
\mathrm{GeV}$. At the tree-level, the model contains three Dirac
neutrinos: one massless, two large with degenerate masses in the
order of the electron mass. At the one-loop level, the left-handed
and right-handed neutrinos obtain Majorana masses $M_{L,R}$ in
orders of $10^{-2}-10^{-3}\ \mathrm{eV}$ and degenerate in
$M_R=-M_L$, while the Dirac masses get a large reduction down to
$\mathrm{eV}$ scale through a finite mass renormalization. In this
model, the contributions of new physics are strongly signified, the
degenerations in the masses and the last hierarchy between the
Majorana and Dirac masses can be completely removed by heavy
particles. All the neutrinos get mass and can fit the data.

\end{abstract}


\maketitle

\section{\label{intro}Introduction}

Until now  the unique evidence supported  physics beyond the
standard model (SM) is nonzero neutrino masses through their
oscillation. The recent experimental results of SuperKamiokande
Collaboration~\cite{superK}, KamLAND~\cite{kam} and SNO~\cite{sno}
confirm that neutrinos have tiny masses and oscillate. This implies
that the standard model must be extended. Among the beyond-SM
extensions, the models based on the $\mathrm{SU}(3)_C\otimes
\mathrm{SU}(3)_L \otimes \mathrm{U}(1)_X$ (3-3-1) gauge group
\cite{ppf,flt} have some intriguing features: First, they can give
partial explanation of the generation number problem. Second,  the
third quark generation has to be different from the first two, so
this leads to the possible explanation of why top quark is
uncharacteristically heavy.

In one of the 3-3-1 models, three lepton triplets are of the form
$(\nu_L,l_L,\nu^c_R)$, where the third members are related to the
right-handed (RH) components of neutrino fields $\nu$. The scalar
sector in this model is minimal with just two Higgs triplets, hence
it has been called the economical 3-3-1 model~\cite{dlns,ponce}. The
general Higgs sector is very simple and consists of three physical
scalars (two neutral and one charged) and eight Goldstone bosons -
the needed number for massive gauge bosons \cite{dls}. The model is
consistent and possesses the key properties: (i) There are three
quite different scales of the vacuum expectation values (VEVs): $u
\sim {\cal O}(1) \ \mathrm{GeV}$, $v \approx 246\ \mathrm{GeV}$, and
$\om \sim {\cal O}(1)\ \mathrm{TeV}$; (ii) There exist two types of
the Yukawa couplings with very different strengths: the
lepton-number conserving (LNC) $h$'s and the lepton-number violating
(LNV) $s$'s satisfying $ s \ll h$. The resulting model yields
interesting physical phenomenologies due to mixings in the Higgs
\cite{dls}, gauge \cite{dlns,dln} and quark \cite{dhhl} sector.

At the tree level, the neutrino spectrum in this model contains
three Dirac fields with one massless and two degenerate in mass
$\sim h^\nu v$, where the Majorana fields $\nu_L$ and $\nu_R$ are
massless. This spectrum is not realistic under the data because
there is only one squared-mass splitting. Moreover, since the
observed neutrino masses are so small, the Dirac mass is unnatural,
and then one must understand what physics gives $h^\nu v \ll h^l v$
- the mass of charged leptons. In contrast to usual cases
\cite{seesaw}, in which the problem can be solved, in this model the
neutrinos including RH ones can only get small masses through
radiative corrections.

The aim of this work is carrying out radiative corrections for the
neutrino masses and gives a possible explanation of why the neutrino
Dirac masses are small. This is not the result of a seesaw
\cite{seesaw}, however, is due to a finite mass renormalization
arising from a very different radiative mechanism. We will show that
the neutrinos can get mass not only from the standard symmetry
breakdown, but also from the electroweak $\mathrm{SU}(3)_L \otimes
\mathrm{U}(1)_X$ breaking associated with spontaneous lepton-number
breaking (SLB), and even through the explicit lepton-number
violating processes of a new physics.

At the one-loop level the Dirac neutrinos can get a large reduction
in mass, the fields $\nu_L$ and $\nu_R$ obtain Majorana masses
$M_{L,R}$. The degeneration of $M_R=-M_L$ is removed by heavy
particles. The total mass spectrum for the neutrinos are therefore
neat and can fit the data.

The rest of this  paper is organized as follows: In Section
\ref{model}, we give a brief review of the economical 3-3-1 model,
the mass mechanisms of neutrinos are represented. Section
\ref{neumass} devotes detailed calculations and analysis of the
neutrino mass spectrum. We summarize our results and make
conclusions in the last section - Sec. \ref{conclus}.

\section{\label{model} A review of the economical 3-3-1 model}

The particle content in this model, which is anomaly free, is given
as follows \bea \psi_{aL} &=& \left(
               \nu_{aL}, l_{aL}, (\nu_{aR})^c
\right)^T \sim (3, -1/3),\hs l_{aR}\sim (1, -1),\hs a = 1, 2, 3,
\crn
 Q_{1L}&=&\left( u_{1L},  d_{1L}, U_L \right)^T\sim
 \left(3,1/3\right),\hs Q_{\al L}=\left(
  d_{\al L},  -u_{\al L},  D_{\al L}
\right)^T\sim (3^*,0),\hs \al=2,3,\crn u_{a
R}&\sim&\left(1,2/3\right),\hs d_{a R} \sim \left(1,-1/3\right),\hs
U_{R}\sim \left(1,2/3\right),\hs D_{\al R} \sim
\left(1,-1/3\right),\eea where the values in the parentheses denote
quantum numbers based on the
$\left(\mbox{SU}(3)_L,\mbox{U}(1)_X\right)$ symmetry. The electric
charge operator in this case takes a form\be
Q=T_3-\fr{1}{\sqrt{3}}T_8+X,\label{eco}\ee where $T_i$
$(i=1,2,...,8)$ and $X$, respectively, stand for $\mbox{SU}(3)_L$
and $\mbox{U}(1)_X$ charges. The electric charges of the exotic
quarks $U$ and $D_\al$ are the same as of the usual quarks, i.e.,
$q_{U}=2/3$, $q_{D_\al}=-1/3$.

The spontaneous symmetry breaking in this model is obtained by two
stages: \be \mathrm{SU}(3)_L\otimes \mathrm{U}(1)_X \rightarrow
\mathrm{SU}(2)_L\otimes\mathrm{U}(1)_Y \rightarrow
\mathrm{U}(1)_Q.\ee The first stage is achieved by a Higgs scalar
triplet with a VEV given by \bea \chi=\left(\chi^0_1, \chi^-_2,
\chi^0_3 \right)^T \sim \left(3,-1/3\right),\hs
\langle\chi\rangle=\fr{1}{\sqrt{2}}\left(u, 0, \om
\right)^T.\label{vevc}\eea The last stage is achieved by another
Higgs scalar triplet needed with the VEV as follows \bea
\phi=\left(\phi^+_1, \phi^0_2, \phi^+_3\right)^T \sim
\left(3,2/3\right)\hs \langle\phi\rangle=\fr{1}{\sqrt{2}}\left(0,v,0
\right)^T.\label{vevp}\eea

The Yukawa interactions which induce masses for the fermions can be
written in the most general form: \be {\mathcal
L}_{\mathrm{Y}}={\mathcal L}_{\mathrm{LNC}} +{\mathcal
L}_{\mathrm{LNV}},\ee in which, each part is defined by \bea
{\mathcal L}_{\mathrm{LNC}}&=&h^U\bar{Q}_{1L}\chi
U_{R}+h^D_{\al\beta}\bar{Q}_{\al L}\chi^* D_{\beta R}\crn
&&+h^l_{ab}\bar{\psi}_{aL}\phi
l_{bR}+h^\nu_{ab}\ep_{pmn}(\bar{\psi}^c_{aL})_p(\psi_{bL})_m(\phi)_n
\crn && +h^d_{a}\bar{Q}_{1 L}\phi d_{a R}+h^u_{\al a}\bar{Q}_{\al
L}\phi^* u_{aR}+ H.c.,\label{y1}\\ {\mathcal
L}_{\mathrm{LNV}}&=&s^u_{a}\bar{Q}_{1L}\chi u_{aR}+s^d_{\al
a}\bar{Q}_{\al L}\chi^* d_{a R}\crn && +s^D_{ \al}\bar{Q}_{1L}\phi
D_{\al R}+s^U_{\al }\bar{Q}_{\al L}\phi^* U_{R}+ H.c.,\label{y2}\eea
where $p$, $m$ and $n$ stand for $\mathrm{SU}(3)_L$ indices.

The VEV $\om$ gives mass for the exotic quarks $U$, $D_\al$ and the
new gauge bosons $Z^{\prime},\ X,\ Y$, while the VEVs $u$ and $v$
give mass for the quarks $u_a,\ d_a$, the leptons $l_a$ and all the
ordinary gauge bosons $Z,\ W$ \cite{dhhl}. In the next sections we
will provide the detailed analysis of neutrino masses. To keep a
consistency with the effective theory, the VEVs in this model have
to satisfy the constraint \be u^2 \ll  v^2 \ll \om^2.
\label{vevcons} \ee
 In addition we can derive $v\approx
v_{\mathrm{weak}}=246\ \mbox{GeV}$ and $|u| \leq2.46\ \mbox{GeV}$
from, the mass of $W$ boson and the $\rho$ parameter \cite{dlns},
respectively. From atomic parity violation in cesium, the bound
for the mass of new natural gauge boson is given by
$M_{Z^{\prime}}>564 \ \mbox{GeV}$ $(\om
> 1400\ \mbox{GeV})$ \cite{dln}. From the analysis on quark
masses, higher values for $\om$ can be required, for example, up to
$10\ \mbox{TeV}$ \cite{dhhl}.

The Yukawa couplings of (\ref{y1}) possess an extra global symmetry
\cite{changlong,tujo} which is not broken by $ v, \omega$ but by
$u$. From these couplings, one can find the following lepton
symmetry $L$ as in Table \ref{lnumber} (only the fields with nonzero
$L$ are listed; all other fields have vanishing $L$). Here $L$ is
broken by $u$ which is behind $L(\chi^0_1)=2$, i.e., $u$ {\it is a
kind of the SLB scale} \cite{major-models}.
\begin{table}[h]
 \caption{\label{lnumber} Nonzero lepton number $L$
 of the model particles.}
\begin{ruledtabular}
\begin{tabular}{lcccccccc}
  Field
&$\nu_{aL}$&$l_{aL,R}$&$\nu^c_{aR}$ & $\chi^0_1$&$\chi^-_2$ &
$\phi^+_3$ & $U_{L,R}$ & $D_{\alpha L,R}$\\
    \hline \\
        $L$ & $1$ & $1$ & $-1$ & $2$&$2$&$-2$&$-2$&$2$
\end{tabular}
 \end{ruledtabular}
\end{table} It is interesting that the exotic quarks also carry the lepton
number (so-called lepton quarks); therefore, this $L$ obviously does
not commute with the gauge symmetry. One can then construct a new
conserved charge $\cal L$ through $L$ by making a linear combination
$L= xT_3 + yT_8 + {\cal L} I$. Applying $L$ on a lepton triplet, the
coefficients will be determined \be L = \fr{4}{\sqrt{3}}T_8 + {\cal
L} I \label{lepn}.\ee Another useful conserved charge $\cal B$ which
is exactly not broken by $u$, $v$ and $\om$ is usual baryon number:
$B ={\cal B} I$. Both the charges $\mathcal{L}$ and $\mathcal{B}$
for the fermion and Higgs multiplets are listed in
Table~\ref{bcharge}.
\begin{table}[h]
\caption{\label{bcharge} ${\cal B}$ and ${\cal L}$ charges of the
model multiplets.} \begin{ruledtabular}
\begin{tabular}{lcccccccccc}
 Multiplet & $\chi$ & $\phi$ & $Q_{1L}$ & $Q_{\al L}$ &
$u_{aR}$&$d_{aR}$ &$U_R$ & $D_{\al R}$ & $\psi_{aL}$ & $l_{aR}$
\\ \hline \\ $\cal B$-charge &$0$ & $ 0  $ &  $\fr 1 3  $ & $\fr 1 3
$& $\fr 1 3  $ &
 $\fr 1 3  $ &  $\fr 1 3  $&  $\fr 1 3  $&
 $0  $& $0$ \\ \hline \\
 $\cal L$-charge &$\fr 4 3$ & $-\fr 2 3  $ &
   $-\fr 2 3  $ & $\fr 2 3  $& 0 & 0 & $-2$& $2$&
 $\fr 1 3  $& $ 1   $
\end{tabular}
 \end{ruledtabular}
\end{table}
Let us note that the Yukawa couplings of (\ref{y2}) conserve
$\mathcal{B}$, however, violate ${\mathcal L}$ with $\pm 2$ units
which implies that these interactions are much smaller than the
first ones \cite{dhhl}: \be s_a^u, \ s_{\al a}^d,\ s_\al^D, \
s_\al^U \ll h^U,\ h_{\al \bet}^D,\ h_a^d,\ h_{\al
a}^u.\label{dkhsyu}\ee

In this model, the most general Higgs potential has very simple form
\bea V(\chi,\phi) &=& \mu_1^2 \chi^\dag \chi + \mu_2^2 \phi^\dag
\phi + \la_1 ( \chi^\dag \chi)^2 + \la_2 ( \phi^\dag \phi)^2\crn & &
+ \la_3 ( \chi^\dag \chi)( \phi^\dag \phi) + \la_4 ( \chi^\dag
\phi)( \phi^\dag \chi). \label{poten} \eea It is noteworthy that
$V(\chi,\phi)$ does not contain trilinear scalar couplings and
conserves both the mentioned global symmetries, this makes the Higgs
potential much simpler and discriminative from the previous ones of
the 3-3-1 models \cite{changlong,tujo,ochoa2}. The non-zero values
of $\chi$ and $\phi$ at the minimum value of $V(\chi,\phi)$ can be
obtained by\bea \chi^+\chi&=&\fr{\lambda_3\mu^2_2
-2\lambda_2\mu^2_1}{4\lambda_1\lambda_2-\lambda^2_3}
\equiv\fr{u^2+\om^2}{2},\label{vev1}\\
\phi^+\phi&=&\fr{\lambda_3\mu^2_1
-2\lambda_1\mu^2_2}{4\lambda_1\lambda_2-\lambda^2_3}
\equiv\fr{v^2}{2}.\label{vev2}\eea Any other choice of $u,\ \om$ for
the vacuum value of $\chi$ satisfying (\ref{vev1}) gives the same
physics because it is related to (\ref{vevc}) by an
$\mbox{SU}(3)_L\otimes \mbox{U}(1)_X$ transformation. It is worth
noting that the assumed $u\neq 0$ is therefore given in a general
case.  This model of course leads to the formation of Majoron
\cite{dls,pheno-331majoron,major-models}. The analysis in \cite{dls}
shows that, after symmetry breaking, there are eight Goldstone
bosons and three physical scalar fields - the SM like $H^0$, the new
neutral $H^0_1$ and the charged bilepton $H^\pm_2$ with the masses:
\be m^2_{H}\simeq\fr{4\la_1\la_2-\la^2_3}{2\la_1}v^2,\hs
m^2_{H_1}\simeq 2\la_1\om^2,\hs
m^2_{H_2}\simeq\fr{\la_4}{2}\om^2.\ee Let us remind the reader that
the couplings $\la_{4,1,2}$ are positive and fixed by the Higgs
boson masses and the $\la_3$, where the last one $\la_3$ is
constrained by $|\la_3|<2\sqrt{\la_1\la_2}$ and gives the splitting
$\Delta m^2_{H}\simeq-[\la^2_3/(2\la_1)]v^2$ from the SM prediction.

In the considering model, the possible different mass-mechanisms for
the neutrinos can be summarized through the three dominant
$\mathrm{SU}(3)_C\otimes \mathrm{SU}(3)_L \otimes
\mathrm{U}(1)_X$-invariant effective operators as follows
\cite{wbg}: \bea O^{\mathrm{LNC}}_{ab}
&=&\bar{\psi}^c_{aL}\psi_{bL}\phi, \label{yka}\\
O^{\mathrm{LNV}}_{ab} &=&(\chi^*\bar{\psi}^c_{aL})(\chi^*\psi_{bL}),\label{heavyparticles}\\
O^{\mathrm{SLB}}_{ab}
&=&(\chi^*\bar{\psi}^c_{aL})(\psi_{bL}\phi\chi),\label{loops}\eea
where the Hermitian adjoint operators are not displayed. It is worth
noting that they are also all the performable operators with the
mass dimensionality $d\leq6$ responsible for the neutrino masses.
The difference among the mass-mechanisms can be verified through the
operators. Both (\ref{yka}) and (\ref{loops}) conserve
$\mathcal{L}$, while (\ref{heavyparticles}) violates this charge
with two units. Since $d(O^{\mathrm{LNC}})=4$ and
$L\langle\phi\rangle=0$, (\ref{yka}) provides only Dirac masses for
the neutrinos which can be obtained at the tree level through the
Yukawa couplings in (\ref{y1}). Since $d(O^{\mathrm{SLB}})=6$ and
$(L\langle\chi\rangle)_p \neq 0$ for $p=1$, vanishes for other
cases, (\ref{loops}) provides both Dirac and Majorana masses for the
neutrinos through radiative corrections mediated by the model
particles. The masses induced by (\ref{yka}) are given by the
standard $\mathrm{SU}(2)_L\otimes \mathrm{U}(1)_Y$ symmetry breaking
via the VEV $v$. However, those by (\ref{loops}) are obtained from
both the stages of $\mathrm{SU}(3)_L\otimes \mathrm{U}(1)_X$
breaking achieved by the VEVs $u,\ \om$ and $v$.

Let us recall that, except the unconcerned LNV couplings of
(\ref{y2}), all the remaining interactions of the model (lepton
Yukawa couplings (\ref{y1}), Higgs self-couplings (\ref{poten}), and
etc.) conserve $\mathcal{L}$, this means that the operator
(\ref{heavyparticles}) cannot be induced mediated by the model
particles. As a fact, the economical 3-3-1 model including the
alternative versions \cite{flt,ppf} are only extensions beyond the
SM in the scales of orders of TeV \cite{dln,ochoa}. Such processes
are therefore expected mediated by heavy particles of an underlined
new physics at a scale $\mathcal{M}$ much greater than $\om$ which
have been followed in various of grand unified theories (GUTs)
\cite{wbg,331GUTs,GUTs}. Thus, in this model the neutrinos can get
mass from three very different sources widely ranging over the mass
scales: $u\sim \mathcal{O}(1)\ \mathrm{GeV}$, $v \approx 246\
\mathrm{GeV}$, $\om \sim \mathcal{O}(1)\ \mathrm{TeV}$, and
$\mathcal{M}\sim \mathcal{O}(10^{16})\ \mathrm{GeV}$.

We remind that, in the former version \cite{flt}, the authors in
\cite{diasalex} have considered operators of the type
(\ref{heavyparticles}), however, under a discrete symmetry
\cite{discr}. As shown by us \cite{dhhl}, the current model is
realistic, and such a discrete symmetry is not needed, because, as a
fact that the model will fail if it enforces. In addition, if such
discrete symmetries are not discarded, the important mass
contributions for the neutrinos mediated by model particles are then
suppressed; for example, in this case the remaining operators
(\ref{yka}) and (\ref{loops}) will be removed. With the only
operator (\ref{heavyparticles}) the three active neutrinos will get
effective zero-masses under a type II seesaw \cite{seesaw} (see
below); however, this operator occupies a particular importance in
this version.

Alternatively, in such model, the authors in \cite{changlong} have
examined two-loop corrections to (\ref{heavyparticles}) by the aid
of explicit LNV Higgs self-couplings, and using a fine-tuning for
the tree-level Dirac masses of (\ref{yka}) down to current values.
However, as mentioned, this is not the case in the considering
model, because our Higgs potential (\ref{poten}) conserves
$\mathcal{L}$. We know that one of the problems of the 3-3-1 model
with RH neutrinos is associated with the Dirac mass term of
neutrinos. In the following, we will show that, if such a
fine-tuning is done so that these terms get small values, then the
mass generation of neutrinos mediated by model particles is not able
to be done or trivially, this is in contradiction with
\cite{changlong}. In the next, the large bare Dirac masses for the
neutrinos, which are as of charged fermions of a natural result from
standard symmetry breaking, will be studied.

\begin{figure}
\includegraphics{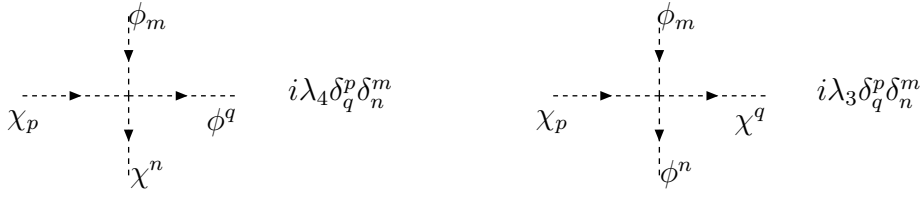}
\caption{\label{figh1} Relevant Higgs self-couplings.}
\end{figure}

\begin{figure}
\includegraphics{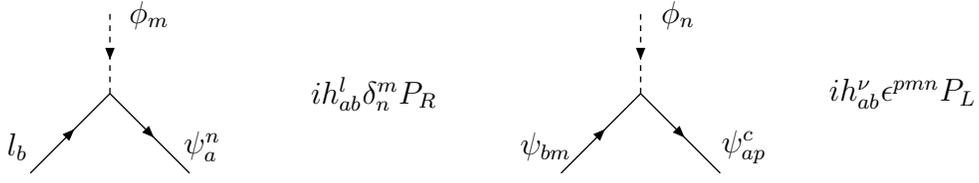}
\caption{\label{figh2} Lepton Yukawa couplings.}
\end{figure}

For the sake of convenience in further reading, we present the
lepton Yukawa couplings and the relevant Higgs self-couplings in
terms represented by Feynman diagrams in Figs. (\ref{figh1}) and
(\ref{figh2}), where the Hermitian adjoint ones are not displayed.

\section{\label{neumass}Neutrino mass matrix}
The operators $O^{\mathrm{LNC}}$, $O^{\mathrm{SLB}}$ and
$O^{\mathrm{LNV}}$ (including their Hermitian adjoint) will provide
the masses for the neutrinos: the first responsible for tree-level
masses, the second for one-loop corrections, and the third for
contributions of heavy particles.

\subsection{\label{treelevel} Tree-level Dirac masses}
From the Yukawa couplings in (\ref{y1}), the tree-level mass
Lagrangian for the neutrinos is obtained by \cite{feynmanrules}\bea
\mathcal{L}^\mathrm{LNC}_\mathrm{mass}&=&
h^\nu_{ab}\bar{\nu}_{aR}\nu_{bL}\langle
\phi^0_2\rangle-h^\nu_{ab}\bar{\nu}^c_{aL}\nu^c_{bR}\langle\phi^0_2\rangle+H.c.\crn
&=& 2\langle\phi^0_2\rangle h^\nu_{ab}\bar{\nu}_{aR}\nu_{bL}+H.c.
=-(M_D)_{ab}\bar{\nu}_{aR}\nu_{bL}+H.c.\crn
&=&-\fr{1}{2}\left(\bar{\nu}^c_{aL},\bar{\nu}_{aR}\right) \left(
\begin{array}{cc}
 0 & (M^T_D)_{ab} \\
(M_D)_{ab} & 0 \\
 \end{array}
\right)\left(
\begin{array}{c}
\nu_{bL} \\
\nu^c_{bR} \\
\end{array}
\right)+H.c. \crn &=& -\fr{1}{2}\bar{X}^c_L M_\nu X_L +H.c., \eea
where $h^\nu_{ab}=-h^\nu_{ba}$ is due to Fermi statistics. The $M_D$
is the mass matrix for the Dirac neutrinos: \bea
(M_D)_{ab}&\equiv&-\sqrt{2}v h^\nu_{ab}=(-M^T_D)_{ab}=\left(
\begin{array}{ccc}
0 & -A & -B \\
A & 0 & -C \\
B & C & 0 \\
\end{array}
\right),\hs
 A,B,C \equiv
\sqrt{2}h^\nu_{e\mu}v,\ \sqrt{2}h^\nu_{e\tau}v,\
\sqrt{2}h^\nu_{\mu\tau}v. \label{dirac1} \eea This mass matrix has
been rewritten in a general basis $X^T_L \equiv(\nu_{eL},\nu_{\mu
L},\nu_{\tau L}, \nu^c_{e R},\nu^c_{\mu R},\nu^c_{\tau R})$: \be
M_\nu \equiv \left(
\begin{array}{cc}
0 & M_D^T \\
M_D & 0 \\
\end{array}
\right).\label{treedirac} \ee

The tree-level neutrino spectrum therefore consists of only Dirac
fermions. Since $h^\nu_{ab}$ is antisymmetric in $a$ and $b$, the
mass matrix $M_D$ gives one neutrino massless and two others
degenerate in mass: $0,\ -m_D,\ m_D$, where $m_D\equiv
(A^2+B^2+C^2)^{1/2}$. This mass spectrum is not realistic under the
data, however, it will be severely changed by the quantum
corrections, the most general mass matrix can then be written as
follows
\bea M_{\nu}=\left(%
\begin{array}{cc}
  M_{L} & M_D^T \\
  M_D & M_{R} \\
\end{array}%
\right),\label{matran}\eea where $M_{L,R}$ (vanish at the
tree-level) and $M_D$ get possible corrections.

If such a tree-level contribution dominates the resulting mass
matrix (after corrections), the model will provide an explanation
about a large splitting either $\Delta m^2_\mathrm{atm} \gg \Delta
m^2_\mathrm{sol}$ or $\Delta m^2_\mathrm{LSND} \gg \Delta
m^2_\mathrm{atm,sol}$ \cite{changlong,pdg}. We then, however, must
need a fine-tuning at the tree-level \cite{changlong} either $m_D
\sim (\Delta m^2_\mathrm{atm})^{1/2}$ $(\sim 5\times10^{-2}\
\mathrm{eV})$ or $m_D \sim (\Delta m^2_\mathrm{LSND})^{1/2}$ $(\sim
\ \mathrm{eV})$ \cite{pdg}. Without lose the generality we can
assume $h^\nu_{e\mu}\sim h^\nu_{e\tau}\sim h^\nu_{\mu\tau}$ which
give us then $h^\nu\sim 10^{-13}\ (\mbox{or}\ 10^{-12})$. The
coupling $h^\nu$ in this case is so small and therefore this
fine-tuning is not natural \cite{pecsmr}. Indeed, as shown below,
since $h^\nu$ enter the dominant corrections from (\ref{loops}) for
$M_{L,R}$, these terms $M_{L,R}$ get very small values which are not
large enough to split the degenerate neutrino masses into a
realistic spectrum. (The largest degenerate splitting in
squared-mass is still much smaller than $\Delta m^2_\mathrm{sol}\sim
8\times10^{-5}\ \mathrm{eV}^2$ \cite{pdg}.) In addition, in this
case the Dirac masses get corrections trivially.

The status of this problem can be changed with the induced operator
(\ref{heavyparticles}) (see below), however, not interested in this
work, because as mentioned the operator (\ref{loops}) that obtains
the contributions of model particles is then discarded. This implies
that the tree-level Dirac mass term for the neutrinos by its
naturalness should be treated as those as of the usual charged
fermions resulted of the standard symmetry breaking, say, $h^\nu\sim
h^{e} \ (\sim 10^{-6})$ \cite{pecsmr}. It turns out that this term
is regarded as a large bare quantity and unphysical. Under the
interactions, they will of course change to physical masses. In the
following we will obtain such {\it finite renormalizations} in the
masses of neutrinos.

\subsection{\label{radia} One-loop level Dirac and Majorana masses}
\begin{figure}
\includegraphics{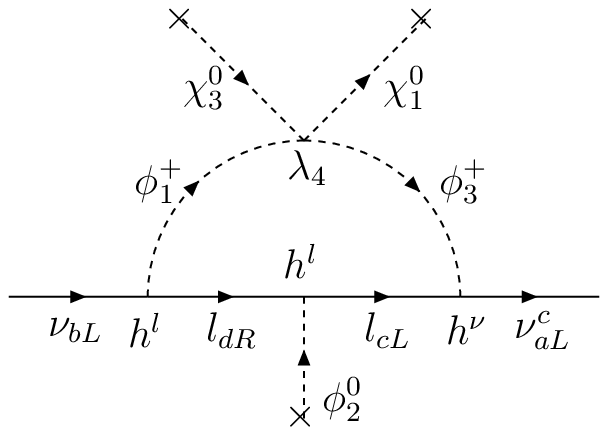}
\includegraphics{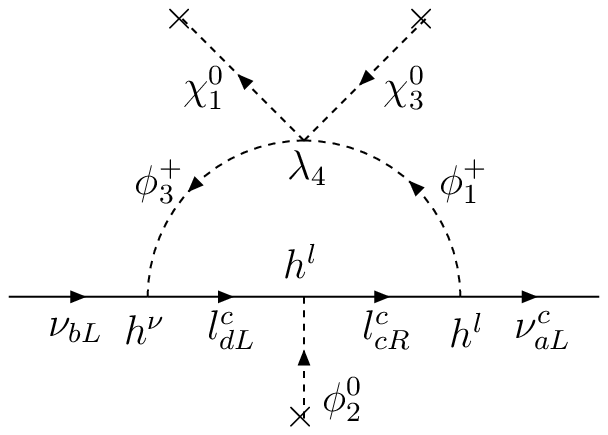}
\caption{\label{hinh12} The one-loop corrections for the mass matrix
$M_L$.}
\end{figure}

\begin{figure}
\includegraphics{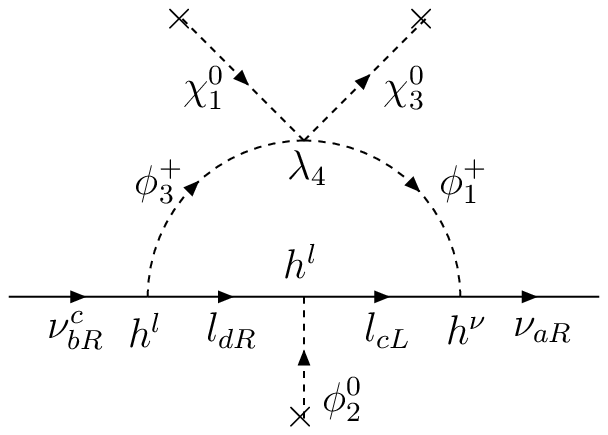}
\includegraphics{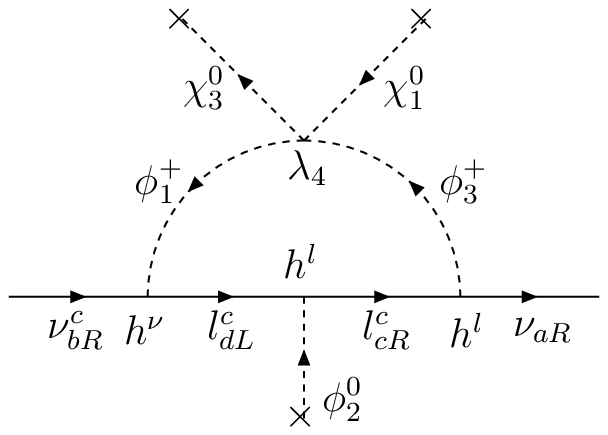}
\caption{\label{hinh34} The one-loop corrections for the mass matrix
$M_R$.}
\end{figure}

\begin{figure}
\includegraphics{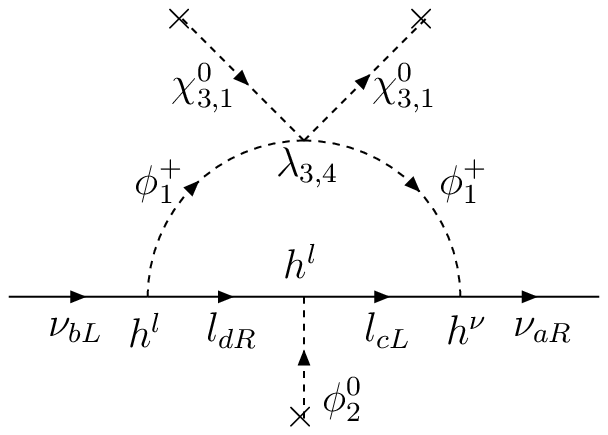}
\includegraphics{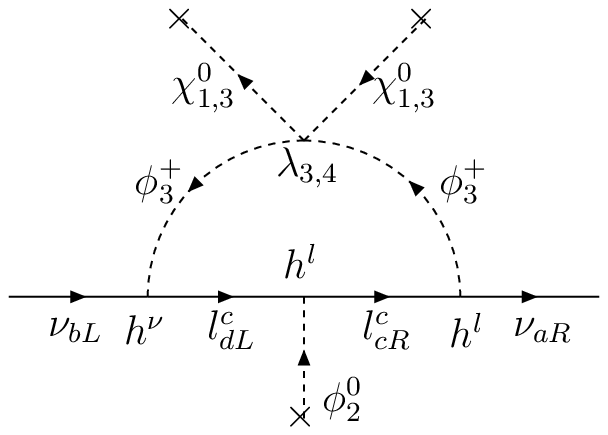}
\caption{\label{hinh56} The one-loop corrections for the mass matrix
$M_D$.}
\end{figure}

The operator (\ref{loops}) and its Hermitian adjoint arise from the
radiative corrections mediated by the model particles, and give
contributions to Majorana and Dirac mass terms $M_L$, $M_R$ and
$M_D$ for the neutrinos. The Yukawa couplings of the leptons in
(\ref{y1}) and the relevant Higgs self-couplings in (\ref{poten})
are explicitly rewritten as follows \bea
\mathcal{L}^\mathrm{lept}_\mathrm{Y}&=&
2h^\nu_{ab}\bar{\nu}^c_{aL}l_{bL}\phi^+_3
-2h^\nu_{ab}\bar{\nu}_{aR}l_{bL}\phi^+_1+h^l_{ab}\bar{\nu}_{aL}l_{bR}\phi^+_1
+h^l_{ab}\bar{\nu}^c_{aR}l_{bR}\phi^+_3+H.c.,\crn
\mathcal{L}^\mathrm{relv}_\mathrm{H}&=&\la_3\phi^-_1\phi^+_1(\chi^{0*}_1\chi^{0}_1
+\chi^{0*}_3\chi^{0}_3)+\la_3\phi^-_3\phi^+_3(\chi^{0*}_1\chi^{0}_1+\chi^{0*}_3\chi^{0}_3)
\crn &+&\la_4\phi^-_1
\phi^+_1\chi^{0*}_1\chi^0_1+\la_4\phi^-_3\phi^+_3\chi^{0*}_3\chi^0_3
+\la_4\phi^-_3\phi^+_1\chi^{0*}_1\chi^0_3 +\la_4\phi^-_1
\phi^+_3\chi^{0*}_3\chi^0_1.\eea The one-loop corrections to the
mass matrices $M_L$ of $\nu_L$, $M_R$ of $\nu_R$ and $M_D$ of $\nu$
are therefore given in Figs. (\ref{hinh12}), (\ref{hinh34}) and
(\ref{hinh56}), respectively.

\subsubsection{Radiative corrections to $M_L$ and $M_R$}

With the Feynman rules at hand \cite{feynmanrules}, $M_L$ is
obtained by \bea
-i(M_L)_{ab}P_L&=&\int\fr{d^4p}{(2\pi)^4}\left(i2h^\nu_{ac}P_L\right)
\fr{i(p\!\!\!/+m_c)}{p^2-m^2_c}\left(ih^l_{cd}\fr{v}{\sqrt{2}}P_R\right)
\fr{i(p\!\!\!/+m_d)}{p^2-m^2_d}\crn
&\times&(ih^{l*}_{bd}P_L)\fr{-1}{(p^2-m^2_{\phi_1})
(p^2-m^2_{\phi_3})}\left(i\la_4\fr{u\om}{2}\right)\crn
&+&\int\fr{d^4p}{(2\pi)^4}\left(ih^{l*}_{ac}P_L\right)
\fr{i(-p\!\!\!/+m_c)}{p^2-m^2_c}\left(ih^l_{dc}\fr{v}{\sqrt{2}}P_R\right)
\fr{i(-p\!\!\!/+m_d)}{p^2-m^2_d}\crn
&\times&(i2h^{\nu}_{bd}P_L)\fr{-1}{(p^2-m^2_{\phi_1})
(p^2-m^2_{\phi_3})}\left(i\la_4\fr{u\om}{2}\right).\label{eq1}\eea
For the sake of simplicity, in the following, we suppose that the
Yukawa coupling of charged leptons $h^l$ is flavor diagonal, thus
$l_c$ and $l_d$ are mass eigenstates respective to the mass
eigenvalues $m_c$ and $m_d$. The equation (\ref{eq1}) becomes then
\bea (M_L)_{ab}=\fr{i\sqrt{2}\la_4u\om}{v} h^\nu_{ab}\left[m^2_b
I(m^2_b,m^2_{\phi_3},m^2_{\phi_1})-m^2_a
I(m^2_a,m^2_{\phi_3},m^2_{\phi_1})\right],\hs (a,b \mbox{ not
summed}),\label{ap2}\eea where the integral $I(a,b,c)$ is given in
Appendix \ref{ap1}.

In the effective approximation (\ref{vevcons}), identifications are
given by $\phi^{\pm}_3 \sim H^{\pm}_2$ and $\phi^{\pm}_1 \sim
G^{\pm}_{W}$ \cite{dls}, where $H^{\pm}_2$ and $G^{\pm}_{W}$ as
above mentioned, are the charged bilepton Higgs boson and the
Goldstone boson associated with $W^\pm$ boson, respectively. For the
masses, we have also $m^2_{\phi_3}\simeq m^2_{H_2}\
(\simeq\fr{\la_4}{2}\om^2)$ and $m^2_{\phi_1}\simeq 0$. Using
(\ref{ketqua}), the integrals are given by \bea
I(m^2_a,m^2_{\phi_3},m^2_{\phi_1})\simeq
-\fr{i}{16\pi^2}\fr{1}{m^2_a-m^2_{H_2}}\left[1-\fr{m^2_{H_2}
}{m^2_a-m^2_{H_2}}\ln\fr{m^2_a}{m^2_{H_2}}\right],\hs
a=e,\mu,\tau.\eea Consequently, the mass matrix (\ref{ap2})
becomes\bea (M_L)_{ab} &\simeq& \fr{\sqrt{2}\la_4 u \om
h^\nu_{ab}}{16\pi^2
v}\left[\fr{m^2_{H_2}(m^2_a-m^2_b)}{(m^2_b-m^2_{H_2})(m^2_a-m^2_{H_2})}+\fr{m^2_a
m^2_{H_2}}{(m^2_a-m^2_{H_2})^2}\ln\fr{m^2_a}{m^2_{H_2}}-\fr{m^2_b
m^2_{H_2}}{(m^2_b-m^2_{H_2})^2}\ln\fr{m^2_b}{m^2_{H_2}}\right]\crn
&\simeq& \fr{\sqrt{2}\la_4 u \om h^\nu_{ab}}{16\pi^2 v
m^2_{H_2}}\left[m^2_a\left(1+\ln\fr{m^2_a}{m^2_{H_2}}\right)-
m^2_b\left(1+\ln\fr{m^2_b}{m^2_{H_2}}\right)\right],\label{tdr}\eea
where the last approximation (\ref{tdr}) is kept in the orders up to
$\mathcal{O}[(m^2_{a,b}/m^2_{H_2})^2]$. Since $m^2_{H_2}\simeq
\fr{\la_4}{2} \om^2$, it is worth noting that the resulting $M_L$ is
not explicitly dependent on $\la_4$, however, proportional to
$t_{\theta}=u/\om$ (the mixing angle between the $W$ boson and the
singly-charged bilepton gauge boson $Y$ \cite{dlns}), $\sqrt{2}v
h^\nu_{ab}$ (the tree-level Dirac mass term of neutrinos), and
$m_{H_2}$ in the logarithm scale. Here the VEV $v\approx
v_{\mathrm{weak}}$, and the charged-lepton masses $m_a$
$(a=e,\mu,\tau)$ have the well-known values. Let us note that $M_L$
is symmetric and has vanishing diagonal elements.

For the corrections to $M_R$, it is easily to check that the
relationship $(M_R)_{ab}=-(M_L)_{ab}$ is exact at the one-loop
level. (This result can be derived from Fig. (\ref{hinh34}) in a
general case without imposing any additional condition on $h^l$,
$h^\nu$, and further.) Combining this result with (\ref{tdr}), the
mass matrices are explicitly rewritten as follows \bea
(M_L)_{ab}=-(M_R)_{ab}\simeq \left(
\begin{array}{ccc}
0 & f & r \\
f & 0 & t \\
r & t & 0 \\
\end{array}\right),\label{majo} \eea where the elements are obtained by
\bea
f&\equiv&\left(\sqrt{2}vh^\nu_{e\mu}\right)\left\{\left(\fr{t_{\theta}}{8\pi^2v^2}\right)
\left[m^2_e\left(1+\ln\fr{m^2_e}{m^2_{H_2}}\right)
-m^2_\mu\left(1+\ln\fr{m^2_\mu}{m^2_{H_2}}\right)\right]\right\},\crn
r&\equiv&\left(\sqrt{2}vh^{\nu}_{e\tau}\right)\left\{\left(\fr{t_{\theta}}{8\pi^2v^2}\right)
\left[m^2_e\left(1+\ln\fr{m^2_e}{m^2_{H_2}}\right)-m^2_\tau
\left(1+\ln\fr{m^2_\tau}{m^2_{H_2}}\right)\right]\right\},\crn
t&\equiv&\left(\sqrt{2}vh^\nu_{\mu\tau}\right)\left\{\left(\fr{t_{\theta}}{8
\pi^2v^2}\right)\left[m^2_\mu
\left(1+\ln\fr{m^2_\mu}{m^2_{H_2}}\right)-m^2_\tau\left(1+\ln\fr{m^2_\tau}
{m^2_{H_2}}\right)\right]\right\}.\label{majo1}\eea

It can be checked that $f,r,t$ are much smaller than those of $M_D$.
To see this, we can take $m_e\simeq 0.51099\ \mathrm{MeV}$, $m_\mu
\simeq 105.65835\ \mathrm{MeV}$, $m_\tau \simeq 1777\ \mathrm{MeV}$,
$v \simeq 246\ \mathrm{GeV}$, $u \simeq 2.46\ \mathrm{GeV}$, $\om
\simeq 3000\ \mathrm{GeV}$, and $m_{H_2}\simeq 700\ \mathrm{GeV}\
(\la_4 \sim 0.11)$ \cite{dlns,dls,dln}, which give us then \be
f\simeq\left(\sqrt{2}vh^\nu_{e\mu}\right)\left(3.18\times
10^{-11}\right), \hs
r\simeq\left(\sqrt{2}vh^{\nu}_{e\tau}\right)\left(5.93\times
10^{-9}\right), \hs
t\simeq\left(\sqrt{2}vh^\nu_{\mu\tau}\right)\left(5.90\times
10^{-9}\right),\label{majo2}\ee where the second factors rescale
negligibly with $\om \sim 1-10\ \mathrm{TeV}$ and $m_{H_{2}}\sim
200-2000\ \mathrm{GeV}$. This thus implies that \be
|M_{L,R}|/|M_D|\sim 10^{-9},\label{cont1}\ee which can be checked
with the help of $|M|\equiv (M^+M)^{1/2}$. In other words, the
constraint is given as follows \be |M_{L,R}|\ll |M_D|.\label{b12}
\ee

With the above results at hand, we can then get the masses by
studying diagonalization of the mass matrix (\ref{matran}), in
which, the submatrices $M_{L,R}$ and $M_D$ satisfying the constraint
(\ref{b12}), are given by (\ref{majo}) and (\ref{dirac1}),
respectively. In calculation, let us note that, since $M_D$ has one
vanishing eigenvalue, $M_\nu$ does not possess the pseudo-Dirac
property in all three generations \cite{kobcsl}, however, is very
close to those because the remaining eigenvalues do. As a fact, we
will see that $M_\nu$ contains a combined framework of the seesaw
\cite{seesaw} and the pseudo-Dirac \cite{pseudo-Dirac}. To get mass,
we can suppose that $h^\nu$ is real, and therefore the matrix $iM_D$
is Hermitian: $(iM_D)^+=iM_D$ (\ref{dirac1}). The Hermitianity for
$M_{L,R}$ is also followed by (\ref{majo}). Because the dominant
matrix is $M_D$ (\ref{b12}), we first diagonalize it by biunitary
transformation \cite{chengli}: \bea
\bar{\nu}_{aR}&=&\bar{\nu}_{iR}(-iU)^+_{ia},\hs
\nu_{bL}= U_{bj}\nu_{jL},\hs (i,j=1,2,3),\label{csl}\\
M_\mathrm{diag}&\equiv& \mathrm{diag}(0,-m_D,m_D)=(-iU)^+ M_D U,\hs
m_D = \sqrt{A^2+B^2+C^2}, \eea where the matrix $U$ is easily
obtained by \bea U=\fr{1}{m_D\sqrt{2(A^2+C^2)}}\left(
\begin{array}{ccc}
C\sqrt{2(A^2+C^2)} & iBC-Am_D & BC-iAm_D \\
-B\sqrt{2(A^2+C^2)} & i(A^2+C^2) & (A^2+C^2) \\
A\sqrt{2(A^2+C^2)} & iAB+Cm_D & AB+iCm_D \\
\end{array}
\right).\eea Resulted by the anti-Hermitianity of $M_D$, it is worth
noting that $M_\nu$ in the case of vanishing $M_{L,R}$
(\ref{treedirac}) is indeed diagonalized by the following unitary
transformation: \bea V=\fr{1}{\sqrt{2}}\left(
\begin{array}{cc}
U & U \\
-iU & iU \\
\end{array}
\right).\eea

A new basis $(\nu_1,\nu_2,...,\nu_6)^T_L \equiv V^+X^T_L$, which is
different from $(\nu_{jL},\nu^c_{iR})^T$ of (\ref{csl}), is
therefore performed. The neutrino mass matrix (\ref{matran}) in this
basis becomes \bea V^+M_\nu V &=& \left(
\begin{array}{cc}
M_{\mathrm{diag}} & \ep \\
\ep & -M_{\mathrm{diag}} \\
\end{array}
\right),\label{ben1}\\
\ep &\equiv& U^+M_L U,\hs \ep^+= \ep,\label{ben3} \eea where the
elements of $\ep$ are obtained by \bea \ep_{11}&=&\ep_{22}=\ep_{33}=0, \label{pp1}\\
\ep_{12}&=&i\ep^*_{13}=\left\{[ABm_D+iC(A^2-B^2+C^2)]f+[(C^2-A^2)m_D+2iABC]r\right.\crn
&&\left.+[iA(A^2 -B^2+C^2)-BCm_D]t\right\}[m^2_D\sqrt{2(A^2 +
C^2)}]^{-1},\label{pp2} \\
\ep_{23}&=&\left\{(A^2+C^2)\left[(Cm_D-iAB)t-(Am_D+iBC)f\right]\right.\crn
&&\left.-\left[B(A^2-C^2)m_D+iAC(A^2+2B^2+C^2)\right]r\right\}
\left[m^2_D(A^2+C^2)\right]^{-1}.\label{pp3}\eea Let us remind the
reader that (\ref{pp1}) is exactly given at the one-loop level
$M_{L}$ (\ref{ap2}) without imposing any approximation on this mass
matrix. Interchanging the positions of component fields in the basis
$(\nu_1,\nu_2,...,\nu_6)^T_L$ by a permutation transformation $P^+
\equiv P_{23}P_{34}$, that is, $\nu_{p} \rightarrow (P^+)_{pq}
\nu_q\ (p,q=1,2,...,6)$ with \be P^+ =\left(
           \begin{array}{cccccc}
             1 & 0 & 0 & 0 & 0 & 0 \\
             0 & 0 & 0 & 1 & 0 & 0 \\
             0 & 1 & 0  & 0 & 0 & 0 \\
             0 & 0 & 1 & 0 & 0 & 0 \\
             0 & 0 & 0 & 0 & 1 & 0 \\
             0 & 0 & 0 & 0 & 0 & 1 \\
           \end{array}
         \right),\ee the mass matrix (\ref{ben1})
can be rewritten as follows \bea P^+(V^+M_\nu V)P &=& \left(
  \begin{array}{cc|cccc}
    0 & 0 & 0 & 0 & \ep_{12} & \ep_{13} \\
    0 & 0 & \ep_{12}& \ep_{13} & 0 & 0 \\ \hline
    0 & \ep_{21} & -m_D & 0 & 0 & \ep_{23} \\
    0 &  \ep_{31} & 0 &  m_D & \ep_{32} & 0 \\
    \ep_{21} & 0 & 0 & \ep_{23} & m_D  & 0 \\
    \ep_{31} & 0 & \ep_{32} & 0 & 0 & -m_D \\
  \end{array}
\right).\label{ben2}\eea

It is worth noting that in (\ref{ben2}) all the off-diagonal
components $|\ep|$ are much smaller than the eigenvalues $|\pm m_D|$
due to the condition (\ref{b12}). The degenerate eigenvalues $0$,
$-m_D$ and $+m_D$ (each twice) are now splitting into three pairs
with six different values, two light and four heavy. The two
neutrinos of first pair resulted by the $0$ splitting have very
small masses as a result of exactly what a seesaw does
\cite{seesaw}, that is, the off-diagonal block contributions to
these masses are suppressed by the large pseudo-Dirac masses of the
lower-right block. The suppression in this case is different from
the usual ones \cite{seesaw} because it needs only the pseudo-Dirac
particles \cite{pseudo-Dirac} with the masses $m_D$ of the
electroweak scale instead of extremely heavy RH Majorana fields, and
that the Dirac masses in those mechanisms are now played by
loop-induced $f,r,t$ (\ref{majo1}) as a result of the SLB $u/\om$.
Therefore, the mass matrix (\ref{ben2}) is effectively decomposed
into $M_{\mathrm{S}}$ for the first pair of light neutrinos
$(\nu_\mathrm{S})$ and $M_{\mathrm{P}}$ for the last two pairs of
heavy pseudo-Dirac neutrinos $(\nu_\mathrm{P})$: \bea
(\nu_1,\nu_4,\nu_2,\nu_3,\nu_5,\nu_6)^T_L \rightarrow
(\nu_\mathrm{S},\nu_\mathrm{P})^T_L=
V^+_{\textmd{eff}}(\nu_1,\nu_4,\nu_2,\nu_3,\nu_5,\nu_6)^T_L,\hs
V^+_{\textmd{eff}}(P^+V^+M_\nu VP)V_{\textmd{eff}}=
\mathrm{diag}\left( M_\mathrm{S},M_\mathrm{P}\right),\eea where
$V_{\textmd{eff}}$, $M_\mathrm{S}$ and $M_\mathrm{P}$ get the
approximations: \bea V_{\textmd{eff}}&\simeq&\left(
\begin{array}{cc}
1 & \mathcal{E} \\
-\mathcal{E}^+ & 1 \\
\end{array}
\right),\hs \mathcal{E}\equiv \left(
  \begin{array}{cccc}
0 & 0 & \ep_{12} & \ep_{13} \\
\ep_{12}& \ep_{13} & 0 & 0 \\
\end{array}\right)\left(\begin{array}{cccc}
-m_D & 0 & 0 & \ep_{23} \\
0 &  m_D & \ep_{32} & 0 \\
0 & \ep_{23} & m_D  & 0 \\
\ep_{32} & 0 & 0 & -m_D \\
\end{array}\right)^{-1},\crn M_{\mathrm{S}}&\simeq& -\mathcal{E}\left(
\begin{array}{cc}
0 & \ep_{21} \\
0 & \ep_{31} \\
\ep_{21}& 0 \\
\ep_{31}& 0 \\
\end{array}
\right),\hs M_{\mathrm{P}}\simeq \left(\begin{array}{cccc}
-m_D & 0 & 0 & \ep_{23} \\
0 &  m_D & \ep_{32} & 0 \\
0 & \ep_{23} & m_D  & 0 \\
\ep_{32} & 0 & 0 & -m_D \\
\end{array}\right).\label{pse}\eea
The mass matrices $M_\mathrm{S}$ and $M_\mathrm{P}$, respectively,
give exact solutions as follows \bea m_{\mathrm{S}\pm}&=&
\pm \fr{2\mathrm{Im}(\ep_{13}\ep_{13}\ep_{32})}{m^2_D-\ep^2_{23}}\simeq
\pm 2\mathrm{Im}\left(\fr{\ep_{13}\ep_{13}\ep_{32}}{m^2_D}\right),\label{p1}\\
m_{\mathrm{P}\pm}&=& -m_D\pm |\ep_{23}|,\hs m_{\mathrm{P'}\pm} = m_D
\pm |\ep_{23}|.\label{p3}\eea In this case, the mixing matrices are
collected into
$(\nu_{\mathrm{S}\pm},\nu_{\mathrm{P}\pm},\nu_{\mathrm{P}'\pm})^T_L
=V^+_{\pm}(\nu_\mathrm{S},\nu_\mathrm{P})^T_L$, where the $V_{\pm}$
is obtained by \be V_{\pm} =\fr{1}{\sqrt{2}}\left(
\begin{array}{cccccc}
1 & -1 & 0 & 0 & 0 & 0 \\
1 & 1 & 0 & 0 & 0 & 0 \\
0 & 0 & \ka & -\ka & 0 & 0 \\
0 & 0 & 0 & 0 & 1 & 1 \\
0 & 0 & 0 & 0 & \ka & -\ka \\
0 & 0 & 1 & 1 & 0 & 0 \\
\end{array}
\right),\hs \ka\equiv \fr{\ep_{23}}{|\ep_{23}|}=\exp(i\arg\ep_{23}).
 \ee It is noted that the degeneration in the Dirac one $|\pm m_D|$ is now
splitting severally.

From (\ref{p3}) we see that the four large pseudo-Dirac masses for
the neutrinos are almost degenerate. In addition, the resulting
spectrum (\ref{p1}), (\ref{p3}) yields two largest squared-mass
splittings, respectively, proportional to $m^2_D$ and $4 m_D
|\ep_{23}|$. Resulted by (\ref{pp3}) and (\ref{majo2}), we can
evaluate $|\ep_{23}|\simeq 3.95\times 10^{-9}\ m_D \ll m_D$ (where
$A\sim B \sim C \sim m_D/\sqrt{3}$ is understood), this therefore
implies that the fine-tuning as mentioned is not realistic because
the splitting $4 m_D |\ep_{23}|$ is still much smaller than $\Delta
m^2_{\mathrm{sol}}$. (In detail, in Table \ref{fine-tuning}, we give
the numerical values of these fine-tunings, where the parameters are
given as before (\ref{majo2}).)
\begin{table}[h]
\caption{\label{fine-tuning} The values for $h^\nu$, elements of
$M_\nu$, and two largest splittings in squared-mass.}
\begin{ruledtabular}
\begin{tabular}{lccccccc}
 Fine-tuning & $h^\nu$ & $\sqrt{2}v h^\nu$ (eV)& $f$ (eV) & $r$ (eV) & $t$ (eV)& $m^2_D\
  (\mathrm{eV}^2)$ & $4 m_D
|\ep_{23}|\ (\mathrm{eV}^2)$ \\
 \hline \\ $m^2_D \sim \Delta m^2_\mathrm{atm}$ &  $8.30\times 10^{-14}$ & $2.88\times 10^{-2}$ &
 $9.18\times 10^{-13}$ & $1.71\times 10^{-10}$ & $ 1.70\times 10^{-10}$
 & $2.50\times 10^{-3}$ & $3.95\times 10^{-11}$\\ \hline \\
 $m^2_D \sim \Delta
m^2_\mathrm{LSND}$ & $1.66\times 10^{-12}$ & $5.77\times 10^{-1}$ &
$1.83\times 10^{-11}$ & $3.43\times 10^{-9}$ & $3.41\times 10^{-9}$
& $1.00$ & $1.58\times 10^{-8}$
\end{tabular}
 \end{ruledtabular}
\end{table}

Similarly, for the two small masses, we can also evaluate
$|m_{\mathrm{S}\pm}|\simeq 4.29\times 10^{-28}\ m_D$. This shows
that the masses $m_{\mathrm{S}\pm}$ are very much smaller than the
splitting $|\ep_{23}|$. This also implies that the two light
neutrinos in this case are hidden for any $m_D$ value of
pseudo-Dirac neutrinos. Let us see the sources of the problem why
these masses are so small: (i) All the elements of left-upper block
of (\ref{ben2}) of the two neutrinos vanish; (ii) In (\ref{p1}) the
resulting masses are proportional to $|\ep|^3/m^2_D$, but not to
$|\ep|^2/m_D$ as expected from (\ref{ben2}). It turns out that this
is due to the antisymmetric of $h^\nu_{ab}$ enforcing on the
tree-level Dirac-mass matrix and the degenerate of $M_R=-M_L$ of the
one-loop level left-handed (LH) and RH Majorana-mass matrices. It
can be easily checked that such degeneration in Majorana masses is
remained up to higher-order radiative corrections as a result of
treating the LH and RH neutrinos in the same gauge triplets with the
model Higgs content; for example, by the aid of (\ref{loops}) the
degeneration is given up to any higher-order loop.

\subsubsection{Radiative corrections to $M_D$}

As mentioned, the mass matrix $M_D$ requires the one-loop
corrections as given in Fig. \ref{hinh56}, and the contributions are
easily obtained as follows \bea
-i(M^{\mathrm{rad}}_D)_{ab}P_L&=&\int\fr{d^4p}{(2\pi)^4}\left(-i2h^\nu_{ac}P_L\right)
\fr{i(p\!\!\!/+m_c)}{p^2-m^2_c}\left(ih^l_{cd}\fr{v}{\sqrt{2}}P_R\right)
\fr{i(p\!\!\!/+m_d)}{p^2-m^2_d}\crn
&\times&(ih^{l*}_{bd}P_L)\fr{-1}{(p^2-m^2_{\phi_1})^2}\left(i\la_3\fr{u^2
+\om^2}{2}+i\la_4\fr{u^2}{2}\right)\crn
&+&\int\fr{d^4p}{(2\pi)^4}\left(ih^{l*}_{ac}P_L\right)
\fr{i(-p\!\!\!/+m_c)}{p^2-m^2_c}\left(ih^l_{dc}\fr{v}{\sqrt{2}}P_R\right)
\fr{i(-p\!\!\!/+m_d)}{p^2-m^2_d}\crn
&\times&(i2h^{\nu}_{bd}P_L)\fr{-1}{(p^2-m^2_{\phi_3})^2}\left(i\la_3\fr{u^2+\om^2}{2}
+i\la_4\fr{\om^2}{2}\right).\label{eq2}\eea We rewrite \bea
(M^{\mathrm{rad}}_D)_{ab}&=&-\fr{i\sqrt{2}h^\nu_{ab}}{v}\left\{\left[\la_3(u^2+\om^2)+\la_4
u^2\right]m^2_b I(m^2_b,m^2_{\phi_1})\right.\crn
&&\left.+\left[\la_3(u^2+\om^2)+\la_4 \om^2\right]m^2_a
I(m^2_a,m^2_{\phi_3})\right\},\hs (a,b \mbox{ not
summed}),\label{eq3}\eea where $I(a,b)$ is given in (\ref{ap4}).
With the help of (\ref{ap3}), the approximation for (\ref{eq3}) is
obtained by\bea
(M^{\mathrm{rad}}_D)_{ab}&\simeq&-\fr{h^\nu_{ab}}{8\sqrt{2}\pi^2
v}\left\{\left[\la_3(u^2+\om^2)+\la_4
u^2\right]+\left[\la_3(u^2+\om^2)+\la_4
\om^2\right]\fr{m^2_a}{m^2_{H_2}} \right\}\crn
&=&-\sqrt{2}h^\nu_{ab}\left(\fr{\la_3\om^2}{16\pi^2
v}\right)\left[1+\left(1+\fr{\la_4}{\la_3}\right)\left(\fr{u^2}{\om^2}+\fr{m^2_a}{m^2_{H_2}}\right)+
\mathcal{O}\left(\fr{u^4}{\om^4},\fr{m^4_{a,b}}{m^4_{H_2}}\right)\right].\label{dirac2}\eea
Because of the constraint (\ref{vevcons}) the higher-order
corrections $\mathcal{O}(\cdot\cdot\cdot)$ can be neglected, thus
$M^{\mathrm{rad}}_D$ is rewritten as follows\be
(M^{\mathrm{rad}}_D)_{ab}=-\sqrt{2}h^\nu_{ab}\left(\fr{\la_3\om^2}{16\pi^2
v}\right)\left(1+\delta_a\right),\hs \delta_a \equiv
\left(1+\fr{\la_4}{\la_3}\right)\left(\fr{u^2}{\om^2}+\fr{m^2_a}{m^2_{H_2}}\right),
\label{epsi}\ee where $\delta_a$ is of course an infinitesimal
coefficient, i.e., $|\delta|\ll 1$. Again, this implies also that if
the fine-tuning is done the resulting Dirac-mass matrix get
trivially. It is due to the fact that the contribution of the term
associated with $\de_a$ in (\ref{epsi}) is then very small and
neglected, the remaining term gives an antisymmetric resulting
Dirac-mass matrix, that is therefore unrealistic under the data.

With this result, it is worth noting that the scale \be
\left|\fr{\la_3\om^2}{16\pi^2 v}\right|\label{sc} \ee of the
radiative Dirac masses (\ref{epsi}) is in the orders of the scale
$v$ of the tree-level Dirac masses (\ref{dirac1}). Indeed, if one
puts $|(\la_3\om^2)/(16 \pi^2 v)| = v$ and takes $|\la_3|\sim
0.1-1$, then $\om \sim 3-10\ \mathrm{TeV}$ as expected in the
constraints \cite{dln,ochoa}. The resulting Dirac-mass matrix which
is combined of (\ref{dirac1}) and (\ref{epsi}) therefore gets two
typical examples of the bounds: (i) $(\la_3\om^2)/(16 \pi^2 v)+v
\sim \mathcal{O}(v) $; (ii) $(\la_3\om^2)/(16 \pi^2 v)+v \sim
\mathcal{O}(0)$. The first case (i) yields that the status on the
masses of neutrinos as given above is remained unchanged and
therefore is also trivial as mentioned. In the last case (ii), the
combination of (\ref{dirac1}) and (\ref{epsi}) gives \be
(M_D)_{ab}=\sqrt{2}h^\nu_{ab}(v\de_a).\label{fmr}\ee It is
interesting that in this case the scale $v$ for the Dirac masses
(\ref{dirac1}) gets naturally a large reduction, and we argue that
this is not a fine-tuning. Because the large radiative mass term in
(\ref{epsi}) is canceled by the tree-level Dirac masses, we mean
this as a finite renormalization in the masses of neutrinos. It is
also noteworthy that, unlike the case of the tree-level mass term
(\ref{dirac1}), the mass matrix (\ref{fmr}) is now {\it
nonantisymmetric in $a$ and $b$ }. Thus all the three eigenvalues of
this matrix are nonzero. Let us recall that in the first case (i)
the zero eigenvalue is, however, retained because the combination of
(\ref{dirac1}) and (\ref{epsi}) is proportional to $h^{\nu}_{ab}v$.

In contrast to ({\ref{cont1}), in this case there is no large
hierarchy between $M_{L,R}$ and $M_D$. To see this explicitly, let
us take the values of the parameters as given before (\ref{majo2}),
thus $\la_3 \simeq -1.06$ and the coefficients $\de_a$ are evaluated
by \be \de_{e}\simeq 6.03\times 10^{-7},\hs\de_{\mu}\simeq
6.23\times 10^{-7},\hs \de_{\tau}\simeq6.28\times
10^{-6}.\label{deta}\ee Hence, we get \be |M_{L,R}|/|M_D| \sim
10^{-2}-10^{-3}.\label{her}\ee With the values given in
(\ref{deta}), the quantities $h^\nu$ and $m_D$ can be evaluated
through the mass term (\ref{fmr}); the neutrino data imply that
$h^\nu$ and $m_D$ are {\it in the orders of $h^e$ and $m_e$} - the
Yukawa coupling and mass of electron, respectively. It is worth
checking that the two largest squared-mass splittings as given
before can be applied on this case of (\ref{her}), and seeing that
they fit {\it naturally} the data.

\subsection{\label{lnv} Mass contributions from heavy particles}

There remain now two questions not yet answered: (i) The
degeneration of $M_R=-M_L$; (ii) The hierarchy of $M_{L,R}$ and
$M_D$ (\ref{her}) can be continuously reduced? As mentioned, we will
prove that the new physics gives us the solution.

The mass Lagrangian for the neutrinos given by the operator
(\ref{heavyparticles}) can be explicitly written as follows \bea
\mathcal{L}^{\mathrm{LNV}}_{\mathrm{mass}}&=&s^\nu_{ab}\mathcal{M}^{-1}
(\langle\chi^*\rangle\bar{\psi}^c_{aL})(\langle\chi^*\rangle\psi_{bL})+H.c.\crn
&=& s^\nu_{ab}\mathcal{M}^{-1}\left(\fr{u}{\sqrt{2}}\bar{\nu}^c_{aL}
+\fr{\om}{\sqrt{2}}\bar{\nu}_{aR}\right)\left(\fr{u}{\sqrt{2}}\nu_{bL}
+\fr{\om}{\sqrt{2}}\nu^c_{bR}\right)+H.c.\crn
&=&-\fr{1}{2}\bar{X}^c_L M^{\mathrm{new}}_{\nu} X_L + H.c.,\eea
where the mass matrix for the neutrinos is obtained by \be
M^{\mathrm{new}}_\nu \equiv-\left(
\begin{array}{cc}
\fr{u^2}{\mathcal{M}}s^\nu & \fr{u\om}{\mathcal{M}}s^\nu \\
\fr{u\om}{\mathcal{M}}s^\nu & \fr{\om^2}{\mathcal{M}}s^\nu\\
\end{array}
\right),\label{heve}\ee in which, the coupling $s^\nu_{ab}$ is
symmetric in $a$ and $b$. Because of the hierarchies $u^2\ll u\om
\ll \om^2$ corresponding of the submatrices $M^{\mathrm{new}}_L$,
$M^{\mathrm{new}}_D$ and $M^{\mathrm{new}}_R$ of (\ref{heve}), the
two questions as mentioned are {\it solved simultaneously}.
Intriguing comparisons between $s^\nu$ and $h^\nu$ are given in
order \ben
\item $h^\nu$ conserves the lepton number; $s^\nu$ violates this
charge.\item $h^\nu$ is antisymmetric and enforcing on the
Dirac-mass matrix; $s^\nu$ is symmetric and breaks this
property.\item $h^\nu$ preserves the degeneration of $M_R=-M_L$;
$s^\nu$ breaks the $M_R=-M_L$.\item A pair of $(s^\nu,h^\nu)$ in the
lepton sector will complete the rule played by the quark couplings
$(s^{q},h^{q})$ \cite{dhhl}.\item $h^\nu$ defines the interactions
in the SM scale $v$; $s^\nu$ gives those in the GUT scale
$\mathcal{M}$.\een

Let us now take the values $\mathcal{M}\simeq 10^{16}\
\mathrm{GeV}$, $\om \simeq 3000\ \mathrm{GeV}$, $u\simeq 2.46\
\mathrm{GeV}$ and $s^\nu\sim \mathcal{O}(1)$ (perhaps smaller), the
submatrices $M^{\mathrm{new}}_L\simeq-6.05\times 10^{-7} s^\nu\
\mathrm{eV}$ and $M^{\mathrm{new}}_D\simeq-7.38\times 10^{-4} s^\nu\
\mathrm{eV}$ can give contributions (to the diagonal components of
$M_L$ and $M_D$, respectively) but very small. It is noteworthy that
the last one $M^{\mathrm{new}}_R \simeq -0.9s^\nu\ \mathrm{eV}$ can
{\it dominate} $M_R$.

To summarize, in this model the neutrino mass matrix is combined
by $M_\nu+M^{\mathrm{new}}_\nu$ where the first term is defined by
(\ref{matran}), and the last term by (\ref{heve}); the submatrices
of $M_\nu$ are given in (\ref{majo}) and (\ref{fmr}),
respectively. Dependence on the strength of the new physics
coupling $s^\nu$, the submatrices of the last term,
$M^{\mathrm{new}}_{L}$ and $M^{\mathrm{new}}_{D}$, are included or
removed.

\section{\label{conclus} Conclusions}

The basic motivation of this work is to study neutrino mass in the
framework of the model based on $\mbox{SU}(3)_C\otimes
\mbox{SU}(3)_L \otimes \mbox{U}(1)_X$ gauge group contained minimal
Higgs sector  with right-handed neutrinos. In this paper, the masses
of neutrinos are given by three different sources widely ranging
over the mass scales including the GUT's and the small VEV $u$ of
spontaneous lepton breaking. We have shown that, at the tree-level,
there are three Dirac neutrinos: one massless and two degenerate
with the masses in the order of the electron mass. At the one-loop
level, a possible framework for the finite renormalization of the
neutrino masses is obtained. The Dirac masses obtain a large
reduction, the Majorana mass types get degenerate in $M_R=-M_L$, all
these masses are in the bound of the data. It is emphasized that the
above degeneration is a consequence of the fact that the left-handed
and right-handed neutrinos, in this model, are in the same gauge
triplets. The new physics including the 3-3-1 model are strongly
signified. The degenerations and the hierarchies among the mass
types are completely removed by heavy particles.

We have shown that the three couplings $\la_{1,2,4}$ of the Higgs
potential are constrained by the scalar masses, the remainder
$\la_3$ is free, negative \cite{dls} and now fixed by the neutrino
masses. This means also that the generation of the neutrino masses
leads to a shift (down) in mass of the Higgs boson from the SM
prediction. In this model, the Goldstone boson $G_X\sim \chi^0_1$ of
the non-Hermitian neutral bilepton gauge boson $X^0$ \cite{dls} is
also a Majoron associated with the neutrino Majorana masses. The
coupling of $HG_{X}G_{X}$ is given at the tree-level \cite{dls} and
will provide a considerable contribution in the invisible modes of
decay of the SM Higgs boson.

The resulting mass matrix for the neutrinos consists of two parts
$M_\nu+M^{\mathrm{new}}_\nu$: the first is mediated by the model
particles, and the last is due to the new physics. Upon the
contributions of $M^{\mathrm{new}}_\nu$, the different realistic
mass textures can be produced such as pseudo-Dirac patterns
associated with the seesaw one are obtained in case of the last term
hidden (neglected). In another scenario, that the bare coupling
$h^{\nu}$ of Dirac masses get higher values, for example, in orders
of $h^{\mu,\tau}$, the VEV $\om$ can be picked up to an enough large
value $(\sim \mathcal{O}(10^4-10^5)\ \mathrm{TeV})$ so that the type
II seesaw spectrum is obtained. Such features deserves further
study.

\section*{Acknowledgments}

P. V. D is grateful to the National Center for Theoretical Sciences
of the National Science Council of the Republic of China for
financial support. He thanks Kingman Cheung and Members of the Focus
Group on Particle Physics at National Tsing Hua University for their
kind help and support. H. N. L. is supported by JSPS grant No:
S-06185, he would also like to thank C. S. Lim and Members of
Department of Physics, Kobe University for warm hospitality during
his visit where this work was completed. The work was also supported
in part by National Council for Natural Sciences of Vietnam.
\\[0.3cm]

\appendix
\section{\label{ap1} Feynman integration}
In this Appendix, we present evaluation of the integral \bea
I(a,b,c)&\equiv&
\int\fr{d^4p}{(2\pi)^4}\fr{p^2}{(p^2-a)^2(p^2-b)(p^2-c)},\label{tichphan}\eea
where $a,b,c>0$ and $I(a,b,c)=I(a,c,b)$ should be noted in use.

In the case of $b\neq c$ and $b,c\neq a$, we introduce a well-known
integral as follows\bea
\int\fr{d^4p}{(2\pi)^4}\fr{1}{(p^2-a)(p^2-b)(p^2-c)}=\fr{-i}{16\pi^2}\left\{\fr{a\ln
a}{(a-b)(a-c)}+\fr{b\ln b}{(b-a)(b-c)}+\fr{c\ln
c}{(c-b)(c-a)}\right\}.\label{interg}\eea Differentiating two sides
of this equation with respect to $a$ we have \bea
\int\fr{d^4p}{(2\pi)^4}\fr{1}{(p^2-a)^2(p^2-b)(p^2-c)}&=&\fr{-i}{16\pi^2}
\left\{\fr{\ln a +1}{(a-b)(a-c)}-\fr{a(2a -b -c)\ln a
}{(a-b)^2(a-c)^2}\right.\crn && \left.+\fr{b\ln
b}{(b-a)^2(b-c)}+\fr{c\ln
c}{(c-a)^2(c-b)}\right\}.\label{intergd}\eea Combining
(\ref{interg}) and (\ref{intergd}) the integral (\ref{tichphan})
becomes \bea I(a,b,c)&=&
\int\fr{d^4p}{(2\pi)^4}\left[\fr{1}{(p^2-a)(p^2-b)(p^2-c)} +
\fr{a}{(p^2-a)^2(p^2-b)(p^2-c)}\right]\crn&=&
\fr{-i}{16\pi^2}\left\{\fr{a(2\ln a +1)}{(a-b)(a-c)} -\fr{a^2(2a -b
-c)\ln a }{(a-b)^2(a-c)^2}+\fr{b^2\ln b}{(b-a)^2(b-c)}+\fr{c^2\ln
c}{(c-a)^2(c-b)}\right\}.\eea

If $a,b\gg c$ or $c\simeq 0 $, we have an approximation as follows
\be I(a,b,c)\simeq
-\fr{i}{16\pi^2}\fr{1}{a-b}\left[1-\fr{b}{a-b}\ln\fr{a}{b}\right].
\label{ketqua}\ee

In the other case with $b=c$ and $b\neq a$, we have also \be
I(a,b)\equiv I(a,b,b)=-\fr{i}{16
\pi^2}\left[\fr{a+b}{(a-b)^2}-\fr{2ab}{(a-b)^3}\ln\fr{a}{b}\right],\label{ap4}\ee
where, also, $I(a,b)=I(b,a)$ should be noted in use.

If $b\gg a$ or $a\simeq 0$, we have the following approximation \be
I(a,b)\simeq-\fr{i}{16\pi^2b}.\label{ap3}\ee

Let us note that the above approximations $aI(a,b,c)$ (or
$bI(a,b,c)$) and $bI(a,b)$ are kept in the orders up to
$\mathcal{O}(c/a,c/b)$ and $\mathcal{O}(a/b)$, respectively.

\end{document}